\pdfoutput=1

\documentclass{bmcart}

\usepackage{graphicx}
\RequirePackage{hyperref}
\usepackage[utf8]{inputenc} 

\startlocaldefs
\endlocaldefs

\begin{document}

\begin{frontmatter}

\begin{fmbox}
\dochead{Research}


\title{Mobile phone data's potential for informing infrastructure planning in developing countries}


\author[
   addressref={aff1,aff2},                   
   corref={aff1},                       
   email={hadrien.salat@orange.com}   
]{\inits{HS}\fnm{Hadrien} \snm{Salat}}
\author[
   addressref={aff2},
   email={zbigniew.smoreda@orange.com}
]{\inits{ZS}\fnm{Zbigniew} \snm{Smoreda}}
\author[
   addressref={aff1},
   email={schlaepfer@arch.ethz.ch}
]{\inits{MS}\fnm{Markus} \snm{Schl\"{a}pfer}}


\address[id=aff1]{
  \orgname{Future Cities Laboratory, Singapore-ETH Centre, ETH Z\"{u}rich}, 
  \street{1 Create Way, CREATE Tower \#06-01},                     %
  \postcode{138602}                                
  \city{Singapore},                              
  \cny{Singapore}                                    
}
\address[id=aff2]{%
  \orgname{Sociology and Economics of Networks and Services department, Orange Labs},
  \street{44 Avenue de la R\'{e}publique},
  \postcode{92320}
  \city{Ch\^{a}tillon},
  \cny{France}
}



\end{fmbox}


\begin{abstractbox}

\begin{abstract} 
High quality census data are not always available in developing countries. Instead, mobile phone data are becoming a trending proxy to evaluate population density, activity and social characteristics. They offer additional advantages for infrastructure planning such as being updated in real-time, including mobility information and recording visitors' activity. We combine various data sets from Senegal to evaluate mobile phone data's potential to replace insufficient census data for infrastructure planning in developing countries. As an applied case, we test their ability at predicting domestic electricity consumption. We show that, contrary to common belief, average mobile phone activity does not correlate well with population density. However, it can provide better electricity consumption estimates than basic census data. More importantly, we use curve and network clustering techniques to enhance the accuracy of the predictions, to recover good population mapping potential and to reduce the collection of required data to substantially smaller samples.
\end{abstract}


\begin{keyword}
\kwd{Mobile phone data}
\kwd{Population mapping}
\kwd{Electricity planning}
\kwd{Developing countries}
\end{keyword}


\end{abstractbox}
%

\end{frontmatter}




\section*{Introduction}

Mobile phone data allow, under certain conditions, to recover a map of the population and can potentially simplify the logistics of census data collection \cite{ratti_landscapees_2006,deville_dynamic_2014,ricciato_estimating_2015,vanhoof_assessing_2018}. This could prove particularly useful in developing countries where such costs cannot be overlooked. However, these approaches have only been validated in developed countries where detailed data is available to train the models and where fine-grained data is comparatively easier to access. Furthermore, a primary objective of population mapping is to inform infrastructure planning. In that respect, mobile phone data have a number of advantages over a simple population count. They represent some notion of intensity of activity, include dynamic real-time usage information and contain mobility patterns. For example, significant results have been obtained for the prediction of short-term population dynamics inside cities \cite{reades_cellular_2007,louail_mobile_2015} and for the prediction of detailed socioeconomic characteristics of users from metadata \cite{blumenstock_predicting_2015,steele_mapping_2017,jahani_improving_2017}. However, these methods once again require large amount of fine-grained data, up to the individual level and its associated privacy concerns, to train the models or may require additional sources of data such as satellite images.

Building on these pioneering studies, we propose new methods that are focused on the specific context of developing countries and on long-term infrastructure planning, and that do not require information about identified individuals. Earlier work has revealed mobile phone data's particular potential for predicting electricity demand \cite{martinez-cesena_using_2015,selvarajoo_urban_2018}. In sub-Saharan Africa, electrification rates remain extremely low, without much optimism for a rapid improvement of the situation \cite{contreras_modedelectrification_2006,bernard_impact_2010,sanoh_local_2012,diouf_initiative_2013}. In 2013 in Senegal, when the last census was collected, the average electrification rate in rural areas was as low as 24\%. Paradoxically, mobile phones have still found their way into the homes of about 75\% of households in these same rural areas. In fact, some studies praise the large coverage achieved by mobile phones in the entire African region \cite{aker_mobile_2010,houngbonon_access_2019}. Our aim is to use the resulting data to reduce the logistic costs of gathering information for infrastructure planning in developing countries. For that purpose, we test the possibility of rebuilding census data from mobile phone data and we evaluate the potential of better predicting electricity demand directly from mobile phone data. We have gathered a bulk of data from Senegal to validate our proposed methods.

Our first important result is that the average mobile phone activity is not necessarily well correlated with population density, an idea that became particularly tempting after the seminal work by Lu et al. who predicted population displacements after a natural disaster from mobile phone data \cite{lu_predictability_2012}. However, we show that it can provide better electricity consumption estimates than basic census data. Finally, we propose a curve and network clustering method that allow to accurately recover census and electricity consumption information from smaller samples, with $r^2$ close to one if one half of the data can be collected. This is a step in the direction proposed by the director of UN Global Pulse, Robert Kirkpatrick, who asserted that ``the next phase in call-records research should be cost–benefit analyses that look at the investment needed to conduct a study, roll out an intervention and appraise the advantages for communities.'' \cite{maxmen_can_2019}.

\section*{Results}

\subsection*{Data}

The population density for each commune in Senegal is given by the 2013 census. There are 552 communes of irregular sizes according to the division provided (created in December 2013), including urban communes (communes de ville and communes d'arrondissement) and rural communes (communaut\'{e}s rurales). The population densities' distribution is close to a narrow Poisson distribution, with an average of 2162 inh./km\textsuperscript{2} and a maximum at 54325 inh./km\textsuperscript{2}. Some mobile phone data were provided by the largest Senegalese telecommunication operator, Sonatel. They contain the number of text messages, number of calls and total length of calls made each hour between each of the operator's 1666 communication towers during the year 2013. To estimate the electricity consumption, we used NOAA's average nighttime lights intensity for the year 2013. The intensity is given as a number between 0 and 63 for each cell of a 30 second arc grid. Since Senegal is close to the equator, this grid is regular and its cells measure about 1km per 1km. This data has been cleaned by NOAA of interference from the moonlight, clouds, etc. to the best of their ability.

Since the nighttime lights grid offers the smallest resolution available, it has been used as the geographical reference. The location of the communication towers is mapped onto the grid and two towers are ``merged'' if they fall in the same cell. After this procedure, 1298 tower emplacements remain. Voronoi cells are then drawn around the towers and the population density is computed from their intersections with the communes, weighted by the proportion of each commune inside the Voronoi cell. The end result is a table containing the population, number of texts, number of calls, call lengths and nighttime light intensity per pixel (i.e. roughly per square kilometre) inside each of the 1298 Voronoi cells.

For the record, we also computed the total value of each variable inside the Voronoi cells instead of their density per pixel. We found consistently better correlations between densities compared to total values and the results for total values are therefore not shown in this paper. In addition, 20 cells with 0 population density were discarded from the analysis and 594 cells with 0 nighttime light intensity and an average of only 47 inhabitants per square kilometre are discarded when logarithms are used. The Voronoi cells are further divided into 511 ``low density cells'' and 173 ``high density cells'' corresponding to a density lower or higher than 5000 inhabitants per pixel to distinguish between mainly rural and mixed or purely urban areas.

\subsection*{Estimations from average values}

The average hourly values of the number of text messages and calls and of the total call length per tower are only moderately correlated with the local population density in Senegal, contradicting in this case the results in \cite{deville_dynamic_2014}. The squared Pearson correlation coefficients ($r^2$) are reported in table~\ref{tab1} (first two cases). The scores are higher when all areas are taken into account compared to the scores when low and high density areas are separated. This suggests that we can only obtain a rough distinction between rural areas and cities.

\begin{table}[h!]
\caption{Nighttime lights correlations ($r^2$) from population density, texts, calls and call length average values. The last three rows are multi-Poisson regression correlations.}\label{tab1}
      \begin{tabular}{lcccc}
        \hline
        Population/Direct     &      & Texts & Calls & Length \\ \hline
        All areas       &      & 0.40  & 0.36  & 0.38   \\
        High density    &      & 0.05  & 0.06  & 0.15   \\
        Low density     &      & 0.30  & 0.29  & 0.26   \\ \hline
        Population/Logs       &      & Texts & Calls & Length \\ \hline
        All areas       &      & 0.76  & 0.74  & 0.74   \\
        High density    &      & 0.10  & 0.10  & 0.11   \\
        Low density     &      & 0.58  & 0.57  & 0.58   \\ \hline
        Night/Direct    & Population & Texts & Calls & Length \\ \hline
        All areas       & 0.44 & 0.39  & 0.46  & 0.47   \\
        High density    & 0.15 & 0.22  & 0.20  & 0.22   \\
        Low density     & 0.51 & 0.29  & 0.44  & 0.45   \\ \hline
        Night/Logs      & Population & Texts & Calls & Length \\ \hline
        All areas       & 0.75 & 0.80  & 0.80  & 0.80   \\
        High density    & 0.13 & 0.24  & 0.18  & 0.21   \\
        Low density     & 0.69 & 0.75  & 0.74  & 0.75   \\ \hline
        Night/Population plus &      & Texts & Calls & Length \\ \hline
        All areas       &      & 0.74  & 0.76  & 0.77   \\
        High density    &      & 0.35  & 0.30  & 0.32   \\
        Low density     &      & 0.62  & 0.70  & 0.72   \\ \hline
      \end{tabular}
\end{table}

The mobile phone variables are also moderately correlated with the nighttime light intensity (see the third case of table~\ref{tab1}). Applying a logarithm to the values significantly improves the overall predictions and the predictions for low-density areas, but not for high-density areas (fourth case in table~\ref{tab1}). On the other hand, population density is also only weakly correlated with the nighttime light intensity. This suggests that the information between population density and mobile phone data may be complementary. Observing that the variables are nearly Poisson distributed, we operate a multi-Poisson regression of the lights from the population density and mobile phone data variables. The results are reported in the bottom three rows of table~\ref{tab1}. We excluded the 594 cells with 0 nighttime light intensity to compute the regression coefficients. We find that the predictions are slightly improved for all types of areas. Finally, we introduce a cap at 63 to the predicted values as a posterior rule to mimic the artificial cap of the real nighttime lights provided by NOAA. Doing this increases the $r^2$ further to 0.81-0.83.
The converse does not hold true however, and predicting the population density from the mobile phone data and the residuals between the mobile phone data and the local nighttime lights intensity does not improve the results. This is shown in the supplementary information (table~S1), together with tests of consistency throughout the year (fig.~S3), and an alternative method to compute regressions from approximate data.

\begin{figure}[h!]
  \includegraphics[]{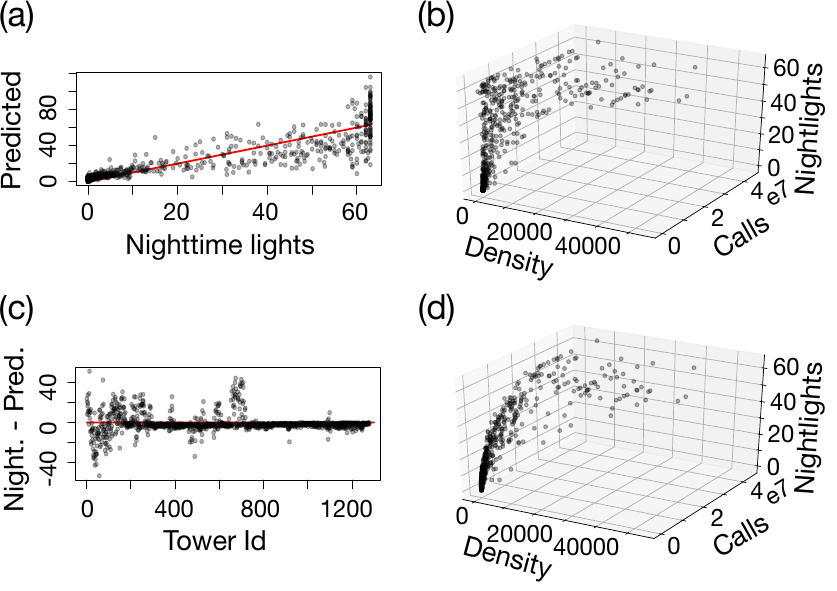}
  \caption{\csentence{Predicted nighttime lights from Poisson regressions of population density and mobile phone data.}\label{fig1}
      (a) Predicted values from total call length against the real values. (c) Difference between the same predicted values and the real values. (b) Real nighttime lights plotted against population density and number of calls. (d) Capped Poisson regression from density and number of calls.}
      \end{figure}

Although encouraging, these results may prove too inaccurate for practical planning in a context of high budget constraints. Fig.~\ref{fig1} illustrates the error between the best predictions, a Poisson regression from the population density and the total number of calls or total call length, and the real values. Panel (a) shows the predicted nighttime lights values plotted against the real values, panel (c) represents the raw difference between the predicted values and the real values and panel (b) represents the real nighttime lights plotted against density and number of calls for comparison with panel (d) showing the capped Poisson regression from density and number of calls.

The population density estimations in particular are not sufficiently reliable. The commodity of replacing census data by mobile data should therefore not be taken for granted. We now propose to tackle the problem from a more different angle by using some of the more hidden information contained in mobile phone data. The aim is threefold: to enhance the general accuracy of the results (possibly at the cost of precision), to bring the density predictions on par with the electricity predictions and to offer an original way to rebuild the census and nighttime lights data from as small a sample as possible.

\subsection*{Estimations from curve and network clustering}

From the mobile data used previously, we compute the average daily, weekly and yearly number of texts, number of calls and total call length (aggregated per hour) curves for each tower site. These 9 types of curves represent the phone usage profiles. They are used to generate distance matrices based on the point-by-point correlation and the standard deviation between the point-by-point distances between two curves. This gives a total of 18 distance clustering matrices. In addition, we attempted to cluster based on curve features inspired by \cite{wang_characteristic-based_2006} including elements such as seasonality, skewness, etc. However, the results were systematically poorer and are not shown here.

\begin{figure}[h!]
\includegraphics[]{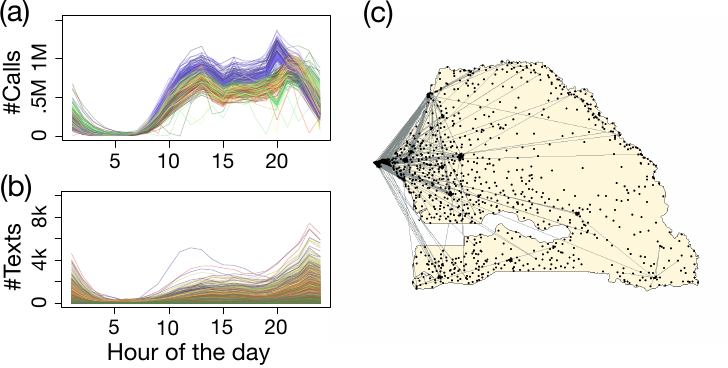}
  \caption{\csentence{Curve profiles and network structure.}\label{fig2}
      (a) Number of calls per hour aggregated at national level for each day of the year. (b) Yearly average of the number of texts per hour of the day sent from each tower. (c) Network structure limited to edges corresponding to at least 2000 text messages sent in January.}
      \end{figure}

In addition, to exploit the characteristic network structure of the data, we transform it into weighted directed graphs averaged over the year. An edge is created between two towers if the activity is above a predefined threshold. We use five thresholds: 0, 60k, 120k, 240k and 480k. The features used for clustering each node are its degree, betweenness and closeness centrality measures (both weighted and unweighted), the ratio of self-loops to the total traffic, the ratio between the number of incoming and outgoing traffic and the average distance travelled by a text message or call. We obtain an additional 15 feature matrices to cluster the curves.
The hourly curves for the number of calls aggregated at national level for each day of the year are represented in fig.~\ref{fig2}(a). There is one colour per month ranging from reds to yellows to greens to blues. The yearly average of number of texts per hour of the day sent from each tower is shown in random colours in fig.~\ref{fig2}(b). The network structure in January limited to edges corresponding to at least 2000 text messages sent is represented in fig.~\ref{fig2}(c).

\begin{figure}[h!]
\includegraphics[width=0.95\linewidth]{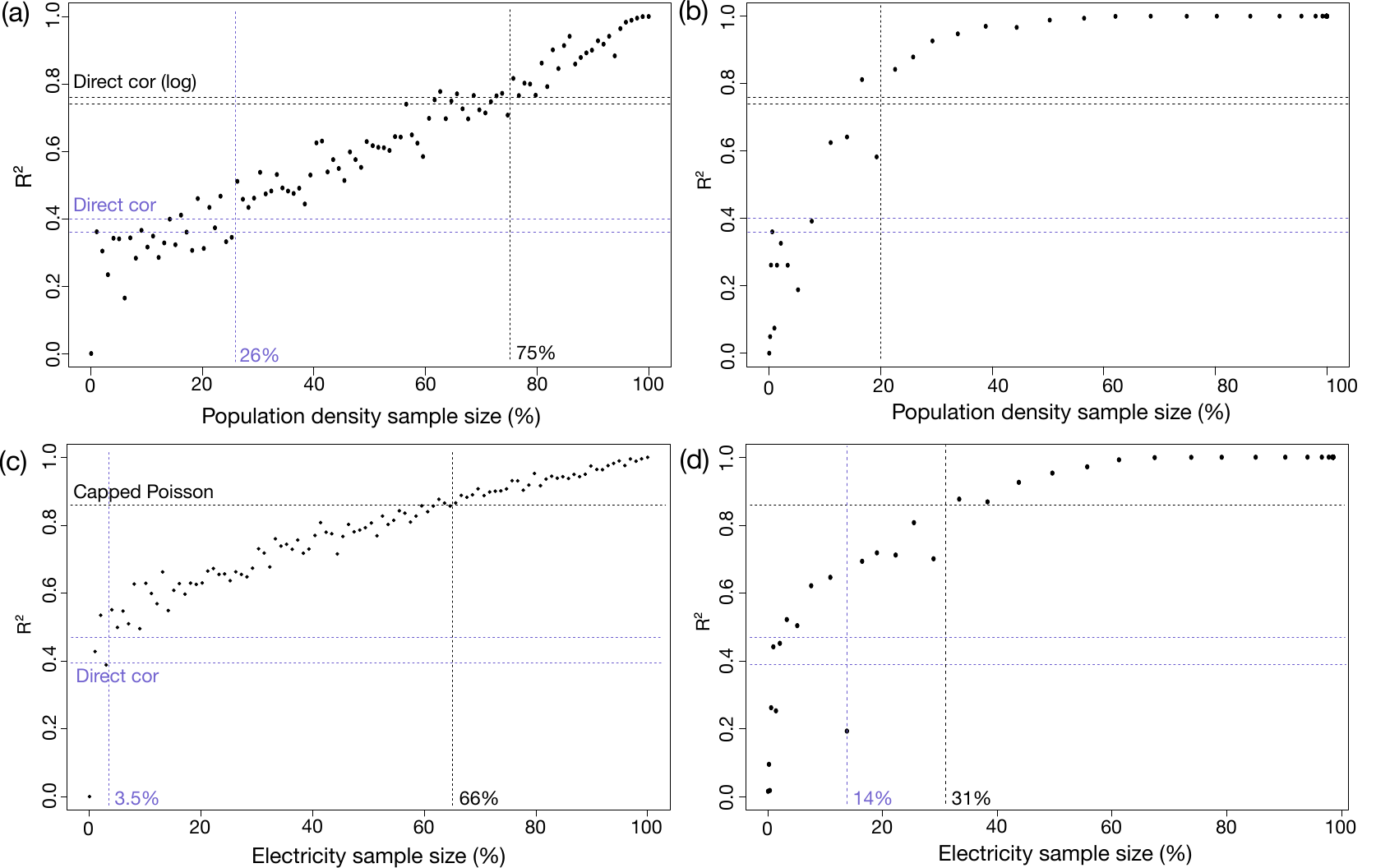}
  \caption{\csentence{Recovering population density and nighttime lights from a sample of reference mobile phone activity curves.}\label{fig3}
      (a,b) Population density. (c,d) nighttime lights intensity. In (a,c), the sample is randomly chosen among all curves. In (b,d), the sample selection is guided by the weekly call length profile clustering tree. The results from the previous methods are shown as horizontal dash lines.}
      \end{figure}

Our objective is to rebuild the original dataset from a sample as small as possible. The first step consists in building a dendrogram from the distance and feature matrices using the \emph{hclust} hierarchical clustering algorithm implemented in \emph{R}. Assume that we know the population or electricity for a number of reference towers. The values of all the other towers are predicted from the proximity of their activity curve to the activity curves of the reference towers. Fig.~\ref{fig3}(a) reports the $r^2$ of the population density distribution predicted from fully random samples of increasing size (expressed as a percentage of the entire distribution). Fig.~\ref{fig3}(b) shows the results when the sampling is guided by the dendrogram produced from the weekly calls activity curves' standard deviation. Panel (c) and (d) repeat the process with electricity distributions. The $r^2$ achieved by the previous methods are indicated as horizontal dash lines. The performance of each sub-method, including the network-based ones, is studied in the supplementary information. Additional figures based on samples guided by some of the other best performing clustering trees are then provided (fig.~S4). Using the tree as a guide has a major impact on the quality of the results. For example, in panel (b), the technique outperforms direct density correlations with samples as small as 20\%, and allows obtaining an $r^2$ of almost 1 for samples as small as 45\%. Since about 75\% of the towers have a low population density, it is indeed unlikely that high density areas would be sufficiently well represented in samples chosen purely randomly.

\begin{figure}[h!]
\includegraphics[width=0.95\linewidth]{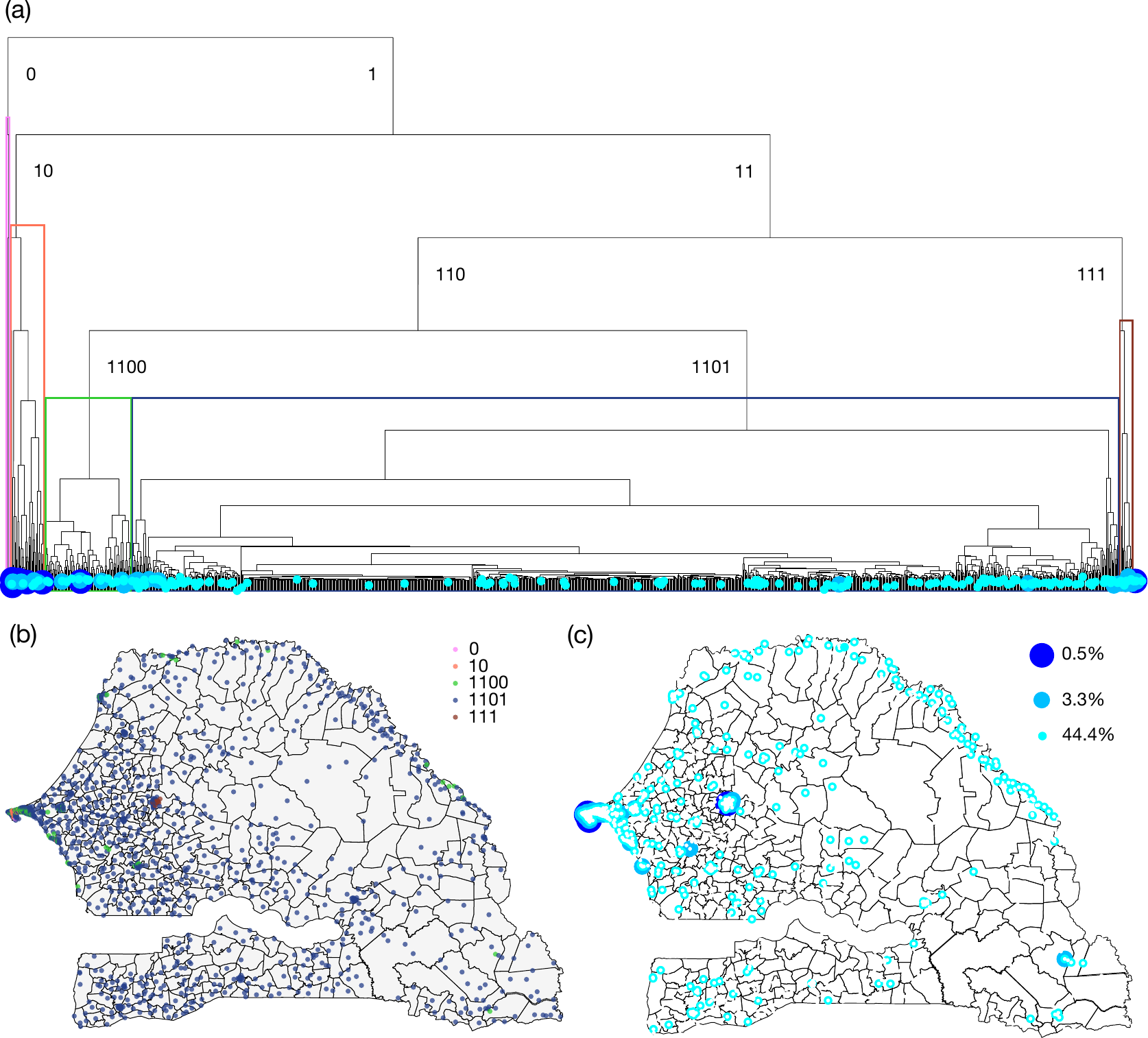}
  \caption{\csentence{Description of the tree cutting process.}\label{fig4}
      (a) Tree corresponding to the weekly call lengths. The branches are identified by a binary number. (b) Locations of the members of five clusters identified on the tree. (c) Locations of one randomly chosen element per branch of the tree cut at depth 3 (dark blue), 7 (medium blue) and 19 (cyan).}
      \end{figure}

By ``guide'', we mean that the tree is first cut at a chosen depth, and then the sample is generated by including one randomly chosen leave per resulting branch. When the selected depth is increased, the sample size is also increased. In particular, the depth can be chosen to match a desired sample size. To illustrate this process, consider the tree corresponding to the weekly call lengths used above. It is shown in fig.~\ref{fig4}(a). Each branch is identified by a binary number counting the number of left turns (indicated by a 0) and right turns (indicated by a 1) that are necessary to reach it while scrolling the tree starting from the top. Five illustrative clusters, evidenced by a colour code, are plotted over a map of Senegal in fig.~\ref{fig4}(b). We can observe that the orange and pink clusters identify most of Dakar, the brown cluster identifies most of Touba (the second biggest city of Senegal), and the green cluster identifies some of the rest of Dakar and some of the locally dense areas, in particular in the middle west region and in the east along the border with Mali. Most of the less dense regions are inside the blue cluster and not yet distinguishable at such a low depth. The process consists in choosing a depth, equivalent to a binary numbers' length, and one random leave in each induced cluster. Examples are given in fig.~\ref{fig4}(c). With a depth of 3 (dark blue), we obtain 7 leaves (or 0.5\% of all towers), with a depth of 7 (medium blue), we obtain 43 leaves (3.3\% of all towers), and with a depth of 19 (cyan), we obtain 576 locations (44.4\% of all towers). Of course, in the event of a branch being reduced to only one leave before the chosen depth is reached, this leave is kept in the sample and the branch is not divided further. There may therefore be fewer elements in the sample than the power of two of the chosen depth.

\begin{figure}[h!]
\includegraphics[]{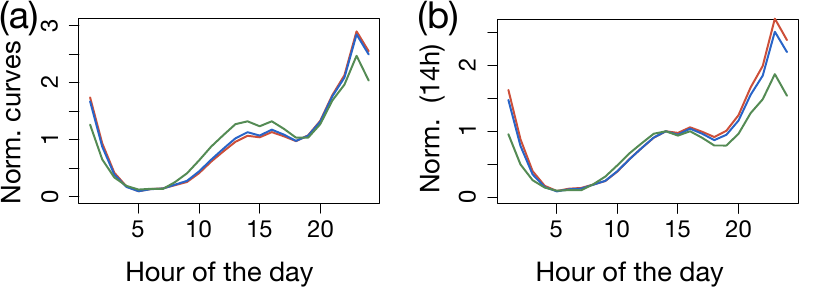}
  \caption{\csentence{Details of the content of the clusters.}\label{fig5}
      (a) Average daily activity curves normalised by the overall daily volume in 3 clusters from the daily texts standard deviation dendrogram. (b) Same average activity curves normalised by the activity at 2 pm.}
      \end{figure}

To get some insights of the hidden functioning of the clustering, we show the average number of text messages sent per hour normalised by the total volume over the day in fig.~\ref{fig5}(a) for a 4 clusters partition of the daily text messages standard deviation tree. Panel (b) shows the same content normalised by the phone traffic at 2 pm. We observe two effects: the green curve corresponding mostly to low density areas is more impacted than the red and blue curves during work to home travel time (4 to 7 pm) and at night. We can indeed hypothesise that the lack of electrification forces people to go to bed earlier in electricity deprived low density areas. Note that one cluster made only of two odd towers has been omitted in the figure.

\section*{Discussion}

We have evidenced two important facts with the estimations from the average values. The first one is that although a positive correlation does exist between mobile phone activity and population density, it is not possible to substitute one for the other in all generality. The second fact is that mobile phone activity contains additional characteristics compared to population density that induce closer correlations between electricity consumption and mobile phone activity than between electricity consumption and population density. As a matter of fact, adding population density to mobile phone activity in the Poisson regressions only increases slightly the fitting of the predictions compared to mobile phone activity alone. We can clearly see in fig.~\ref{fig1}(d) that most of the shape of the nighttime light predictions is due to the higher mobile phone exponent.

With the clustering methodology, we have introduced new perspectives. It is now possible to enforce hard cluster selection rules to ensure that the results are accurate, albeit at the potential expense of precision. More importantly, we have shown that an entire census can be reconstructed from a substantially smaller sample of carefully selected locations with the help of the clustering trees. There are two main practical applications to this: a reduction in data collection costs to about one third in the best case scenarios and the possibility to keep tracking the changes in the population distribution between two census surveys. This technique could complement for example the approach by Lai et. al. which shares the same aim, but uses direct regressions from average values only \cite{lai_exploring_2019}. Note that contrary to previous methods, the clustering relies solely on mobile phone data, without requiring a population training dataset. In addition, it can accommodate external information known \textit{a priori}. It should be possible for example to use satellite data to subset potential reference locations into obviously low or obviously high density areas.

Methodologically speaking, the clean Pareto and linear fits of the second method using average values (fig.~S1) may be worth exploring further. In addition, it appears from the results that the distance matrices based on the standard deviation of the point-wise distances between the curves yield consistently better results than those based on direct correlations. Similarly, it seems that the curve based clustering is more reliable than the network based clustering. However, the network based clustering might prove easier to refine since other more relevant features could be added or ways to integrate over all activity thresholds instead of a discrete selection could be thought of. As mentioned in the data section, we also tested using counts directly instead of spatial densities for all variables (e.g. population, number of calls). The results were significantly poorer, showing that high mobile phone activity and electricity demand might result more from the diversity of intricate activity induced by high population densities rather than just from the raw number of inhabitants. Similar conclusions have been drawn for industrial efficiency \cite{oclery_path_2016} for example.


Finally, the reliance on proprietary data is an obstacle. Although Sonatel, the operator curating the mobile phone data, is the market leader, its market coverage is not uniform over the entire country. Obtaining detailed geographical market share data was not possible at this time. As a result, the quality of some of the correlations may be underestimated in our analysis.


\begin{backmatter}

\section*{Availability of data and materials}
  The census data for the year 2013 in Senegal can be directly accessed through \href{http://www.ansd.sn/index.php?option=com_content&view=article&id=134&Itemid=262}{the official website}. The nighttime lights data can be accessed through The nighttime lights data can be accessed through \href{https://ngdc.noaa.gov/eog/dmsp/downloadV4composites.html}{NOAA's open database}. The mobile phone data at Voronoi level and aggregated over the year are available as part of the supplementary material (Additional file 1). The identity of the callers has been removed and the exact location of the communication towers has been slightly modified for confidentiality reasons. In addition, a time series containing the number of calls per hour for the month of January is also available as part of the supplementary material (Additional file 2). To obtain the dataset over the entire year, one would need to contact Sonatel directly. The analysis was performed using \emph{R}. The terminology used for the clustering methodology is coherent with Murphy's book \cite{murphy_machine_2012}.

\section*{Competing interests}
  The authors declare that they have no competing interests.
  
\section*{Funding}
  H.S. was supported by the Orange Labs-Sonatel-ETH Singapore SEC Research Agreement No.~H11283. M.S. acknowledges the Future Cities Laboratory at the Singapore-ETH Centre, which was established collaboratively between ETH Zurich and Singapore's National Research Foundation (FI~370074016) under its Campus for Research Excellence and Technological Enterprise Programme. 

\section*{Author's contributions}
  H.S., Z.S. and M.S. designed research; H.S. performed research and wrote the paper; Z.S. curated the data.

\section*{Acknowledgements}
  The authors acknowledge Aike Steentoft for his guidance in choosing an adequate methodology to cluster the mobile phone activity curves.


\bibliographystyle{bmc-mathphys} 
\bibliography{clustering}      



\section*{Additional Files}
  \subsection*{Additional file 1 --- Additional methods and figures}
    Alternative method to estimate nightlights intensity from approximate data, validation of the performance of the clustering process and additional figures.

  \subsection*{Additional file 2 --- Mobile phone data at Voronoi level and aggregated over the year}
    The table contains an id of the Voronoi cell, the longitude and latitude coordinates of each Voronoi centre (slightly modified for privacy reasons), the average population density and nighttime light intensity per km\textsuperscript{2} inside the cell, and the number of text messages, calls and total call length per km\textsuperscript{2} for each cell.

  \subsection*{Additional file 3 --- Time series of outgoing calls for each Voronoi cell in January}
    The table contains a Voronoi cell id, a time stamp for each hour of the month and the number of outgoing calls during this hour in the cell.

\end{backmatter}
\end{document}


\begin{frontmatter}

\begin{fmbox}
\dochead{Supplementary information}

\title{Mobile phone data's potential for informing infrastructure planning in developing countries}

\author[
   addressref={aff1},                   
   email={hadrien.salat@orange.com}   
]{\inits{HS}\fnm{Hadrien} \snm{Salat}}
\author[
   email={zbigniew.smoreda@orange.com}
]{\inits{ZS}\fnm{Zbigniew} \snm{Smoreda}}
\author[
   email={schlaepfer@arch.ethz.ch}
]{\inits{MS}\fnm{Markus} \snm{Schl\"{a}pfer}}

\address[id=aff1]{
  \orgname{Future Cities Laboratory, Singapore-ETH Centre, ETH Z\"{u}rich}, 
  \street{1 Create Way, CREATE Tower \#06-01},                     %
  \postcode{138602}                                
  \city{Singapore},                              
  \cny{Singapore}                                    
}

\end{fmbox}
\end{frontmatter}


\section*{Alternative method to estimate nightlights intensity from approximate data}

The practical potential of the regressions in the estimations from average values section remains limited by the requirement of precise training data that may not be available in the context of developing countries. We therefore compare the quality of nighttime lights predictions from population density and mobile phone data using an alternative method that does not require precise data. The population density for each tower is binned into 30 categories and an exponent function is computed to solve the simple equations
\begin{eqnarray}
NIGHT_i &= &DENS_i\alpha(DENS_i),\\
NIGHT_i &= &TEXTS_i\beta(DENS_i),
\end{eqnarray}
where i is used to indicate the values for the bin indexed i. For comparison, we also bin the number of texts into 30 categories and solve the equation
\begin{equation}
NIGHT_i=TEXTS_i\gamma(TEXTS_i).
\end{equation}
The functions $\alpha$, $\beta$ and $\gamma$ are deduced from the binned values and are represented by the red and blue regression curves and lines in fig.~S1. In the third case, the linear (red) and Pareto (blue) fits yield an identical fitting $r^2$ (0.925 vs. 0.923). When the estimated functions are applied to the full data, we find nighttime lights predictions with an $r^2$ of 0.77 for density binned by density, 0.76 for texts binned by density, 0.81 for texts binned by texts with a Pareto fit and 0.86 for texts binned by texts with a linear fit. Remarkably, the well behaved $\alpha$ function allows better fits from the population density than the direct correlations. The similarly well behaved $\gamma$ function allows predictive quality close to the Poisson regressions. More importantly, using the mobile phone activity alone once again outperforms the predictions from population density. The errors generated by the predictions of the last method can be seen in Panel (d) of fig.~S1.

The impact of varying the number of bins (and why setting it to 30 is reasonable) is shown in fig.~S1. Note that the first bin (containing negligible densities) has been discarded to enhance the fittings. For similar reasons, two abnormally high values of text messages were omitted in the third case.

\section*{Validation of the performance of the clustering process}

Assuming that the full data distribution is known beforehand, we define cluster purity conditions and scroll through the dendrogram by splitting branches when they do not meet the specified conditions. The results for some conditions over all 33 trees are shown in table S2. The conditions are formulated as``the cluster must contain at least $x_1\%$ of towers with a value between $y_0$ and $y_1$ or at least $x_2\%$ of towers with a value between $y_1$ and $y_2$, etc.'' The different thresholds for different bins can prove useful if the size of the bins is not homogeneous. We aim at fulfilling the conditions in as few clusters as possible. Once the process has been operated over all trees, a final prediction is computed for each tower. The expected value inside each cluster the tower belongs to (that is one per tree) weighted by the cluster's purity are averaged. We can obtain high $r^2$ with hard conditions at the cost of a high number of clusters (column 1 of table~S2) or a very low number of clusters per tree with softer conditions at the cost of less precision (column 2 of table~S2).


\begin{backmatter}

\section*{Figures}
  \begin{figure}[h!]
  \includegraphics[]{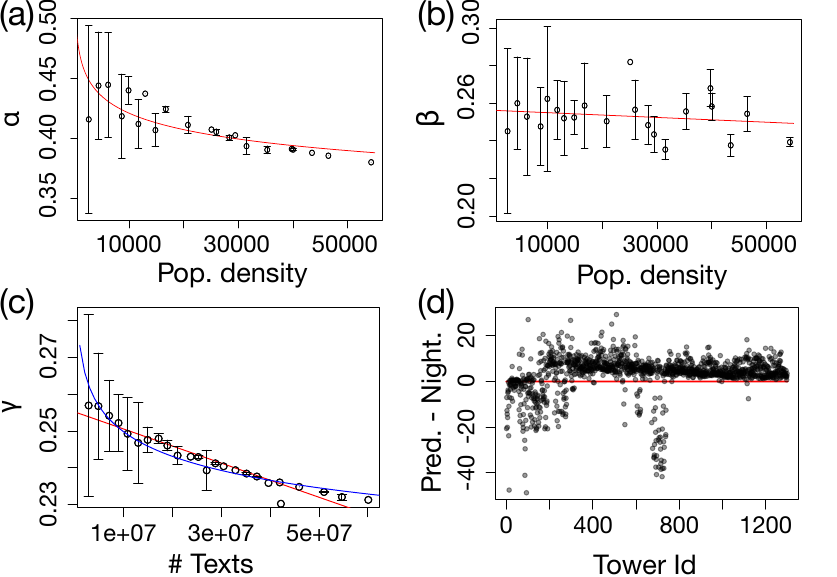}
  \caption{\csentence{Exponent distributions for nighttime lights predictions from binned population density and binned texts messages.}\label{figS1}
       (a) $\alpha$ in eq. (1). (b) $\beta$ in eq. (2). (c) $\gamma$ in eq. (3). (d) Difference between the predicted values and the real values by Voronoi id in case (c) (red linear fitting).}
      \end{figure}

\begin{figure}[h!]
\includegraphics[]{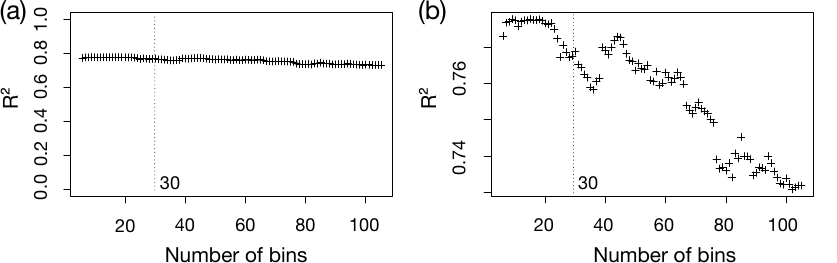}
  \caption{\csentence{Impact of the number of bins on the reliability of the varying coefficient method.}\label{figS2}
       The number of bins starts at 6. The scores correspond to the $\alpha$ coefficient of nighttime lights estimations from density binned by density. (b) is a zoom of (a).}
      \end{figure}
      
\begin{figure}[h!]
\includegraphics[]{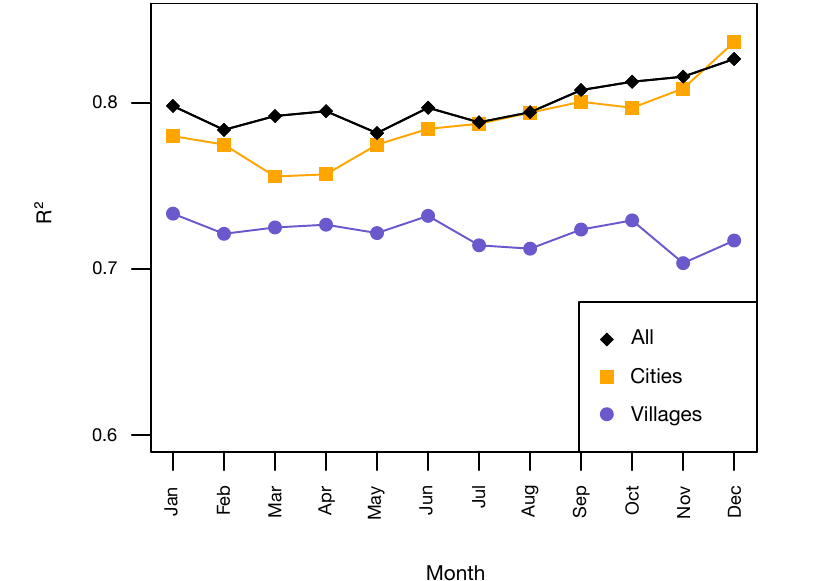}
  \caption{\csentence{Monthly correlations between text messages and nighttime lights.}\label{figS3}
       ``Cities'' refers to higher density areas ($>5000$ inh./km\textsuperscript{2}) and ``Villages'' to lower density areas ($<5000$ inh./km\textsuperscript{2}). We observe a slight increase of the overall scores as months pass by. This can be explained by an also slight increase in the number of towers being activated.}
      \end{figure}
      
\begin{figure}[h!]
\includegraphics[width=\linewidth]{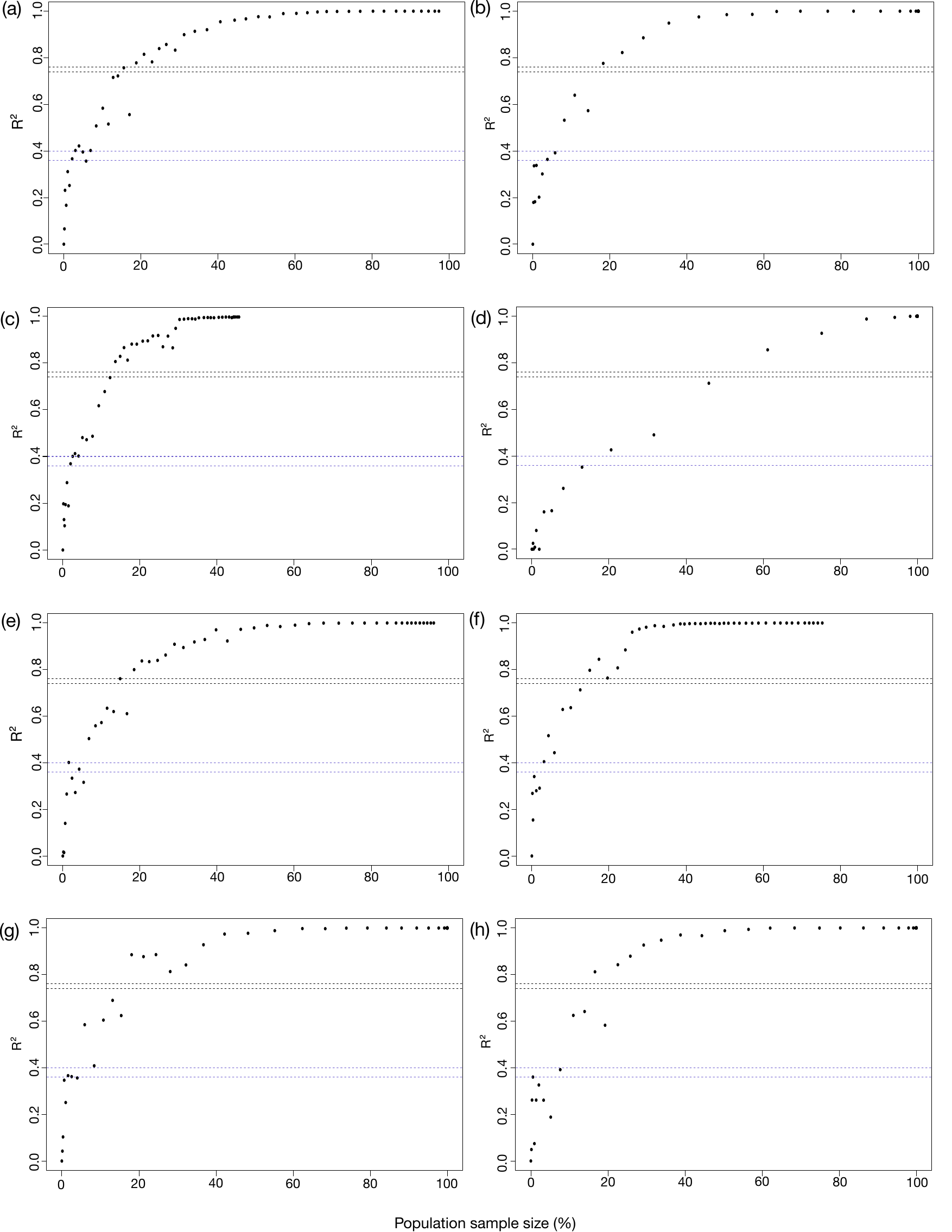}
  \caption{\csentence{Additional results of density estimations from dendrogram guided sampling.}\label{figS4}
       (a) Yearly / Calls / Sd. (b) Daily / Length / Sd. (c) Yearly / Texts / Sd. (d) Daily / Calls / Cor. (e) Yearly / Length / Sd. (f) Weekly/ Texts / Sd. (g) Weekly / Calls / Sd. (h) Weekly / Length / Sd.}
      \end{figure}

\section*{Tables}
\begin{table}[h!]
\renewcommand\thetable{S1}
\caption{Density estimations from mobile phone data and residuals with nighttime lights data.}\label{tabS1}
      \begin{tabular}{lccc}
        \hline
        Residuals    & Texts & Calls & Length \\ \hline
        All areas    & 0.50  & 0.51  & 0.50   \\
        High density & 0.25  & 0.27  & 0.26   \\
        Low density  & 0.20  & 0.19  & 0.19   \\ \hline
      \end{tabular}
\end{table}

\begin{table}[h!]
\renewcommand\thetable{S2}
\caption{ Number of clusters necessary to fulfil the required cluster purity conditions. D, W and Y mean daily, weekly or yearly profiles; T, C and L mean text messages, number of calls and call length; Sd and Cor mean that the standard deviation or correlation method was used; 0, 1, 2, 3, 4 in the Network description indicate the threshold used to subdivide the network (see main text). The thresholds are expressed in percent. Only the upper bounds of the bins are indicated.}\label{tabS2}
\begin{tabular}{lrrrr}
                          & \multicolumn{3}{c}{Population density}                 & Electricity    \\ 
            Bins          & (500,5k,20k,55k) & (500,5k,20k,55k) & (1k,10k,40k,55k) & (10,30,50,63)  \\
            Thresholds    & (82,58,58,58)    & (80,50,50,50)    & (89,65,65,70)    & (88,60,60,60)  \\
\hline
			Curve DTSd    & 117              & 35               & 151              & 119            \\
			Curve DTCor   & 159              & 132              & 234              & 211            \\
			Curve DCSd    & 101              & 27               & 153              & 127            \\
			Curve DCCor   & 111              & 94               & 143              & 129            \\
			Curve DLSd    & 112              & 23               & 176              & 174            \\
			Curve DLCor   & 108              & 71               & 220              & 197            \\
			Curve WTSd    & 129              & 35               & 150              & 114            \\
			Curve WTCor   & 183              & 102              & 270              & 177            \\
			Curve WCSd    & 67               & 56               & 171              & 154            \\
			Curve WCCor   & 142              & 105              & 156              & 135            \\
			Curve WLSd    & 73               & 37               & 145              & 149            \\
			Curve WLCor   & 126              & 99               & 165              & 161            \\
			Curve YTSd    & 126              & 117              & 120              & 94             \\
			Curve YTCor   & 226              & 176              & 274              & 221            \\
			Curve YCSd    & 98               & 28               & 125              & 212            \\
			Curve YCCor   & 250              & 187              & 231              & 208            \\
			Curve YLSd    & 89               & 77               & 156              & 171            \\
			Curve YLCor   & 248              & 68               & 269              & 241            \\
\hline
			Curve $r^2$   & 0.864            & 0.776            & 0.904            & 0.925          \\
\hline
            Network T0    & 177              & 129              & 192              & 158            \\
            Network T1    & 107              & 81               & 90               & 93             \\
            Network T2    & 69               & 48               & 217              & 237            \\
            Network T3    & 206              & 183              & 211              & 155            \\
            Network T4    & 160              & 114              & 1198             & 1081           \\
            Network C0    & 118              & 85               & 139              & 96             \\
            Network C1    & 199              & 147              & 192              & 191            \\
            Network C2    & 65               & 44               & 112              & 91             \\
            Network C3    & 155              & 123              & 464              & 461            \\
            Network C4    & 304              & 269              & 1189             & 1076           \\
            Network L0    & 126              & 109              & 164              & 141            \\
            Network L1    & 170              & 129              & 140              & 183            \\
            Network L2    & 141              & 83               & 165              & 196            \\
            Network L3    & 103              & 33               & 106              & 130            \\
            Network L4    & 164              & 135              & 163              & 191            \\
\hline
            Network $r^2$ & 0.902            & 0.817            & 0.928            & 0.933          \\
\hline
\end{tabular}
\end{table}

\end{backmatter}